\documentstyle[12pt]{article}\title{Higher Order Polarizabilities of the Proton}
\author{B.R. Holstein\\
Institut f\"{u}r Kernphysik\\
Forschungszentrum J\"{u}lich\\
D-52425 J\"{u}lich, Germany\\
and\\
Department of Physics and Astronomy\\
University of Massachusetts\\
Amherst, MA  01003\\
and\\
D. Drechsel, B. Pasquini, M. Vanderhaeghen\\
Institut f\"{u}r Kernphysik\\
Johannes Gutenberg Universit\"{a}t\\
D-55099 Mainz, Germany}
\begin{document}
\begin{titlepage}
\maketitle
\begin{abstract}
Compton scattering results are used to probe proton
structure via measurement of higher order polarizabilities.
Values for $\alpha_{E2}^p,\beta_{E2}^p,\alpha_{E\nu}^p,$
$\beta_{E\nu}^p$ 
determined via dispersion relations are compared to predictions based
upon chiral symmetry and from the constituent quark 
model. Extensions to spin-polarizabilities are also discussed.
\end{abstract}
\end{titlepage}
\section{Introduction}
Recently the availability of high intensity 
electron facilities and tagged photon 
beams has allowed proton structure to be probed by means of Compton
scattering~\cite{Nat96}.
In the case of photons with 
wavelength much larger than the size of the target, 
only the overall charge is resolvable.  
Then, to lowest order the effective Hamiltonian is  
\begin{equation}
H_{eff}^{(0)}={(\vec{p}-e\vec{A})^2\over 2M},
\end{equation}
and the spin-averaged amplitude for 
Compton scattering on the proton is given simply by the 
familiar Thomson form
\begin{equation}
{\rm Amp}^{(0)}=
-{e^2\over M}\hat{\epsilon}_1\cdot\hat{\epsilon}_2, \label{eq:1}
\end{equation}
where $e,M$ represent the proton charge, mass and
$\hat{\epsilon}_1,\hat{\epsilon}_2$ and
$k_1^\mu=(\omega,\vec{k}_1),{k}_2^\mu=(\omega,\vec{k}_2)$ 
specify the polarization
vectors and four-momenta of the initial,final photons respectively.  At higher 
energies (and shorter wavelengths) the structure of the 
system begins to be observable.  The corresponding effective Compton
scattering Hamiltonian must be quadratic in the vector potential 
and be gauge invariant, so
it must be written in terms of the electric and magnetic fields.
It must also be a rotational scalar and invariant 
under parity and time reversal
transformations.  Consequently, the simplest form is~\cite{pedag}
\begin{equation}
H_{eff}^{(2)}=-{1\over 2}4\pi\alpha_E^p\vec{E}^2-{1\over 2}4\pi
\beta_M^p\vec{H}^2,
\end{equation}
and with the difinitions of the electric and magnetic dipole moments 
\begin{equation}
\vec{p}=-{\delta H_{eff}^{(2)}\over \delta\vec{E}}=4\pi\alpha_E^p\vec{E},\quad
\vec{\mu}=-{\delta H_{eff}^{(2)}\over \delta\vec{H}}=4\pi\beta_M^p\vec{H}
\end{equation}
we recognize $\alpha_E^p,\beta_M^p$ as the electric, magnetic polarizabilities 
respectively, which  measure the response of the proton to  
quasistatic electric and magnetizing fields.  The corresponding 
${\cal O}(\omega^2)$ Compton scattering amplitude becomes
\begin{equation}
{\rm Amp}^{(2)}=\hat{\epsilon}_1\cdot\hat{\epsilon}_2\left({-e^2\over
M}+\omega^2 \; 4\pi\alpha_E^p \right)
+\hat{\epsilon}_1\times\hat{k}_1\cdot\hat{\epsilon}_2
\times\hat{k}_2\;\omega^2 4\pi\beta_M^p+\; {\cal O}(\omega^4) 
\end{equation}
and the resultant differential scattering cross section is 
\begin{eqnarray}
{d\sigma\over d\Omega}&=&\left({\alpha\over M}\right)^2
\left[{1\over 2}
(1+\cos^2\theta)\right.\nonumber\\
&-&\left.{ M\omega^2\over \alpha}\left({1\over 2}
(\alpha_E^p+\beta_M^p)(1+\cos\theta)^2+{1\over
2}(\alpha_E^p-\beta_M^p)(1-\cos\theta)^2\right)+\ldots\right],
\nonumber\\
\quad
\end{eqnarray}
where $\alpha=e^2/4\pi$ is the fine structure constant.
Thus by measurement of the differential Compton scattering cross
section one can extract the electric and magnetic polarizabilities,
provided 
\begin{itemize}
\item[i)] the energy is large enough that such terms are
significant with respect to the Thomson contribution, but
\item[ii)]  not so large
that higher order effects begin to dominate.
\end{itemize}
This extraction via the $\gamma p\rightarrow \gamma p$ 
reaction has been accomplished using measurements 
in the energy regime 50 MeV$ <\omega<$100
MeV, yielding~\cite{Nat96}
\begin{equation}
\alpha_E^p=(12.1\pm 0.8\pm 0.5)\times
10^{-4}\,\,{\rm fm}^3;\qquad\beta_M^p=(2.1\mp 0.8\mp 0.5)\times
10^{-4}\,\,{\rm fm}^3 \, .\label{eq:yy}
\end{equation}
Note that in practice one generally uses the results of
unitarity and the validity of the forward ($t$=0) scattering dispersion
relation, which yields the Baldin sum rule~\cite{bal}
\begin{eqnarray}
\alpha+\beta & = & 14.2 \,\pm\, 0.5 \;\;\;
({\rm{Ref.}}~\cite{Damashek}) \nonumber\\
             & = & 13.69 \,\pm\, 0.14 \;\;\; ({\rm{Ref.}}~\cite{BGM})\ .
\label{eq1.5}
\end{eqnarray}
as a constraint, since the uncertainty associated with the integral
over the photo-absorption cross section $\sigma_{\rm tot}(\omega)$ is
smaller than that associated with the polarizability measurements. 
 
On the theoretical side, at the crudest level, we observe that the 
size of $\alpha_E^p$ reveals the feature that the nucleon is strongly bound.
Indeed for the hydrogen atom the electric polarizability is of the order 
of the atomic volume~\cite{merz}
\begin{equation}
\alpha_E^{\rm H-atom}={27\over 8\pi}\times({4\over 3}\pi a_0^3),
\end{equation}  
where $a_0=1/m_e\alpha$ is the Bohr radius.  On the 
other hand, Eq. (\ref{eq:yy}) 
shows that for the proton 
\begin{equation}
\alpha_E^p\sim 4\times 10^{-4}\times({4\over 3}\pi<r_p^2>^{3\over 2}) \, .
\end{equation} 
More quantitative investigations generally involve one of two techniques.  
The first involves use of a nonrelativisitic 
constituent quark picture of the proton
and the quantum mechanical sum rule~\cite{Fr75}
\begin{equation}
\alpha_E^p={1\over 3M}<0|\sum_{i=1}^3e_i(\vec{r}_i-\vec{R}_{cm})^2|0>
+2\sum_{n\neq 0}{|<n|\sum_{i=1}^3e_i(\vec{r}_i-\vec{R}_{cm})_z|0>|^2
\over E_n-E_0}, \label{eq:aa}
\end{equation}
where $e_i,\vec{r}_i$ denotes the charge, position of the $i$th 
constituent quark and $|0>$ represents the ground state.
In this case the simple harmonic oscillator model of nucleon structure is 
found to be somewhat too simplistic, since when the oscillator frequency
is fitted to the charge radius via
\begin{equation}
\omega_0={\sqrt{3}\over M<r_p^2>}\approx 180\,\,{\rm MeV}\label{eq:lll}
\end{equation}
the predicted size of the polarizability 
\begin{equation}
\alpha_E^p={2\alpha M\over 9}<r_p^2>^2\approx 35\times 10^{-4}\,\,{\rm fm}^3
\end{equation}
is a factor of three or so too large.   

The failure here is associated with the low value of the
oscillator frequency given by Eq. (\ref{eq:lll}), and use of a more
realistic excitation energy $\omega_0\simeq 300$ MeV yields a value
in the right ballpark.  However, the real solution to this problem
requires going beyond the simple constituent quark picture of the proton
to consider meson cloud structure~\cite{ww}---{\it i.e.} 
a proper treatment of the pionic degrees of 
freedom---and suggests the efficacy of the second 
approach---heavy baryon chiral perturbation theory (HB$\chi$PT)~\cite{BKM95}.  
Using this technique, one finds at ${\cal
O}(p^3)$ in the chiral expansion~\cite{BKKM92}
\begin{equation}
\alpha_E^p=10K_p=12.7\times 10^{-4}\,\,{\rm fm}^3,\quad
\beta_M^p=K_p=1.3\times 10^{-4}\,\,{\rm fm}^3
\end{equation}
where $K_p=\alpha g_A^2/192\pi F_\pi^2m_\pi$.  Here $g_A\simeq 1.266$
is the axial coupling constant in neutron beta decay and $F_\pi\simeq
92.4$ MeV  is the pion decay constant. 
This ${\cal O}(p^3)$ calculation represents only the leading result for 
$\alpha_E^p$,$\beta_M^p$ in HB$\chi$PT but gets the
qualitative features of the polarizabilities right and even agrees
with experiment! The results diverge as $1/m_\pi$ in the chiral limit, 
giving support to the idea that at these low energies the photon interacts
primarily  with the long-range pion cloud of the nucleon.  Of course, 
one must include higher order terms in order to properly judge the
convergence behavior of the series, and such a calculation at ${\cal O}(p^4)$ 
has been performed by Bernard, Kaiser,
Schmidt and Mei\ss ner (BKSM)~\cite{BKSM93}.  At this order counterterms are
required, which were
estimated by BKSM by treating higher resonances---including $\Delta$(1232)---as very heavy
with respect to the nucleon, yielding
\begin{equation}
\alpha_E^p=(10.5\pm2.0)\times
10^{-4}\,\,{\rm fm}^3;\quad\beta_M^p=(3.5\pm 3.6)\times
10^{-4}\,\,{\rm fm}^3
\end{equation}
where the uncertainty is associated with the counterterm contribution
from the $\Delta(1232)$ and from $K,\eta$ loop effects. 

An alternative tack has been pursued by Hemmert et al., who have developed a
chiral expansion---the small scale or ``$\epsilon$''-expansion---wherein 
the $\Delta(1232)$ is included as an {\it explicit} degree of freedom
and which involves taking $\Delta\equiv M_\Delta-M_N$ as an additional
``small'' parameter~\cite{hem}. 
In this approach, one finds new contributions to the ${\cal
O}(p^3)$ predictions~\cite{gmr1}
\begin{eqnarray}
\delta \alpha_E^p&=&{L_p\over 6}\left({9\Delta\over \Delta^2-m_\pi^2}
+{\Delta^2-10m_\pi^2\over (\Delta^2-m_\pi^2)^{3\over 2}}\ln R\right)\nonumber\\
\delta\beta_M^p&=&{8\over 9}{b_1^2\alpha\over M^2\Delta}+
{L_p\over 6}{1\over (\Delta^2-m_\pi^2)^{1\over 2}}\ln R
\end{eqnarray}
where $L_p=g_{\pi N\Delta}^2\alpha/9\pi^2F_\pi^2$ with $g_{\pi N\Delta}$ being
the $\pi N\Delta$ coupling constant,
$b_1$ the corresponding coupling for radiative $\Delta(1232)$
decay, and
\begin{equation}
R={\Delta\over m_\pi}+\sqrt{{\Delta^2\over m_\pi^2}-1}\, .
\end{equation} 
From the experimentally obtained size of the $\Delta\rightarrow N\pi$
and $\Delta\rightarrow N\gamma$ widths, one determines
$g_{\pi N\Delta}=1.05\pm 0.2$, $b_1=-1.93\pm 0.1$.  Use of these numbers
then results in an increase
in the predicted electric polarizability of about 30\% and takes us away from
experimental agreement at ${\cal O}(\epsilon^3)$.  However, BKSM have shown that there exists a sizable
negative ${\cal O}(p^4)$ $N\pi$ loop contribution which tends to
cancel this discrepancy.  

With respect to the magnetic polarizability, the
simple quark model {\it does} provide a basic understanding.   
The prediction~\cite{Fr75}
\begin{eqnarray}
\beta_M^p&=&-{1\over 2M}(\sum_ie_i(\vec{r}_i-\vec{R}_{cm})^2>
-{1\over 6}<\sum_ie_i^2(\vec{r}_i-\vec{R}_{cm})^2/m_i>\nonumber\\
&+&2\sum_{n\neq 0}{|<n|\sum_i{e_i\over 2m_i}\sigma_{iz}|0>|^2\over E_n-E_0}
\end{eqnarray}
involves a substantial diamagnetic recoil contribution
\begin{equation}
{}^{\rm dia}\beta_M^p=-10.2\times 10^{-4}\,\,{\rm fm}^3 
\end{equation}
which, when added to the large paramagnetic pole contribution due to the 
$\Delta(1232)$~\cite{M9}
\begin{equation}
{}^{\Delta}\beta_{M}^p=+12\times 10^{-4}\,\,{\rm fm}^3,
\end{equation}
yields results in basic agreement with the experimental findings.  
It is clear then that proper inclusion of the $\Delta(1232)$ degrees of
freedom is essential. 

The above summary is intended only as a brief review of the subject
and is not presumed to represent a substitute for more
detailed discussions such as found in Ref.~\cite{lvov}.  However, it
does reveal how important structure information can be obtained via
measurement of the
static polarizabilities.  The purpose of the
present paper is to ask whether it is possible to use Compton scattering 
in order to provide additional proton structure information via the use of 
{\it higher-order} polarizabilities.  Specifically, 
in the next section we define and generate theoretical predictions for
four new  
spin-averaged polarizabilities which arise at ${\cal O}(\omega^4)$ 
in the expansion
of the Compton scattering amplitude.  Then in section 3 we show how 
such quantities
can be extracted from existing experimental data using fixed-t 
dispersion relations
and confront the values obtained thereby with theoretical expectations.
In section 4 we extend our discussion to the case of spin-polarizabilities, and
we conclude with a brief chapter summarizing our findings. 

\section{Quadratic polarizabilities}

As outlined above, the electric and magnetic polarizabilities arise as 
${\cal O}(\omega^2)$ corrections to the lowest order (Thomson) scattering
amplitude.  If one extends the analysis to consider spin-averaged 
${\cal O}(\omega^4)$ terms, then four new structures are possible
which obey the
requirements of gauge, P, and T invariance.  These can be written in the 
form~\cite{rev}
\begin{equation}
H_{eff}^{(4)}=-{1\over 2}4\pi\alpha_{E\nu}^p\dot{\vec{E}}^2
-{1\over 2}4\pi\beta_{M\nu}^p\dot{\vec{H}}^2
-{1\over 12}4\pi\alpha_{E2}^pE_{ij}^2
-{1\over 12}4\pi\beta_{M2}^p H_{ij}^2\label{eq:mmm}
\end{equation} 
where 
\begin{equation}
E_{ij}={1\over 2}(\nabla_iE_j+\nabla_jE_i),\quad H_{ij}={1\over 2}(\nabla_iH_j
+\nabla_jH_i)
\end{equation}
denote electric and magnetizing field gradients. The physical meaning of the
new terms is clear from their definition---Eq. (\ref{eq:mmm}). The quantities
$\alpha_{E\nu}^p,
\beta_{M\nu}^p$ represent dispersive corrections to the lowest order static 
polarizabilities $\alpha_E,\beta_M$ and describe the response of the system to
time dependent fields via (in frequency space)
\begin{eqnarray}
\vec{p}(\omega)&=&4\pi\alpha_E^p(\omega)\vec{E}(\omega)=4\pi(\alpha_E^p
+\alpha_{E\nu}^p\omega^2+\ldots)\vec{E}(\omega)\nonumber\\
\vec{\mu}(\omega)&=&4\pi\beta_M^p(\omega)\vec{H}(\omega)
=4\pi(\beta_M^p+\beta_{M\nu}^p\omega^2+\ldots)
\vec{H}(\omega)\, .
\end{eqnarray}   
The parameters $\alpha_{E2}^p,\beta_{M2}^p$, on the other hand, represent 
quadrupole polarizabilities and measure the electric, magnetic quadrupole
moments induced in a system in the presence of an applied field gradient via
\begin{equation}
Q_{ij}={\delta H_{eff}^{(4)}\over \delta E_{ij}}={1\over
6}4\pi\alpha_{E2}^pE_{ij},
\quad M_{ij}={\delta H_{eff}^{(4)}\over \delta H_{ij}}
={1\over 6}4\pi\beta_{M2}^p H_{ij}
\end{equation}
where
\begin{equation}
Q_{ij}=<p|\sum_ke_k\left[3(\vec{r}_k-\vec{R}_{cm})_i(\vec{r}_k-\vec{R}_{cm})_j-
\delta_{ij}(\vec{r}_k-\vec{R}_{cm})^2\right]|p>
\end{equation}
indicates the proton electric quadrupole moment and $M_{ij}$ is its
magnetic analog.

As in the case of the ordinary polarizabilities one can attempt to predict the
size of these quantities in two somewhat orthogonal ways.  
For example, using the sum rules 
\begin{equation}
\alpha_{E\nu}^p=2\alpha\sum_{n\neq 0}{|z_{n0}|^2\over (E_n-E_0)^3},\quad
\alpha_{E2}^p={\alpha\over 2}\sum_{n\neq 0}{|(Q_{zz})_{n0}|^2\over E_n-E_0}
\label{eq:klm}
\end{equation}
and the simple oscillator picture, one finds the predictions
\begin{equation}
\alpha_{E\nu}^p={2\alpha M^3\over
81}<r_p^2>^4,\quad\alpha_{E2}^p={\alpha M\over 9}<r_p^2>^3\label{eq:nnn}
\end{equation}
and similarly one can generate predictions for the corresponding
magnetic quantities.  However, there is no reason to suspect that this
picture should yield any better results here than in the case of the
ordinary polarizabilities.

On the other hand, 
one can also predict the quadratic polarizabilities within 
heavy baryon chiral perturbation theory using either the ${\cal O}(p^3)$ or
${\cal O}(\epsilon^3)$ expansion.  In the former case one finds~\cite{gmr}
\begin{equation}
\alpha_{E\nu}^p={9\over 10}{K_p\over m_\pi^2},\quad
\beta_{M\nu}^p={7\over 5}{K_p\over m_\pi^2},\quad\alpha_{E2}^p=
{42\over 5}{K_p\over m_\pi^2},\quad\beta_{M2}^p=-{18\over 5}{K_p\over m_\pi^2}
\label{eq:cde}
\end{equation} 
while in the latter the values given in Eq. (\ref{eq:cde}) 
are augmented by the terms
\begin{eqnarray}
\delta\alpha_{E\nu}^p&=&-{1\over 180}{L_p\over m_\pi^2}\left({\Delta\over 
(\Delta^2-m_\pi^2)^3}
(29\Delta^4-143\Delta^2m_\pi^2-231m_\pi^4)\right.\nonumber\\
&-&\left.{3m_\pi^2(\Delta^4
-107\Delta^2m_\pi^2-9m_\pi^4)
\over (\Delta^2-m_\pi^2)^{7\over 2}}\ln R\right)\nonumber\\
\delta\beta_{M\nu}^p&=&{8\over 9}{b_1^2\alpha\over M^2\Delta^3}+{1\over 360}
{L_p\over m_\pi^2}
\left({54\Delta^3-144\Delta m_\pi^2\over (\Delta^2-m_\pi^2)^2}
+{3(2\Delta^2m_\pi^2+28m_\pi^4)\over (\Delta^2-m_\pi^2)^{5\over 2}}\ln R
\right)\nonumber\\
\delta\alpha_{E2}^p&=&{1\over 30}{L_p\over m_\pi^2} \left({22\Delta^3-82\Delta 
m_\pi^2\over (\Delta^2-m_\pi^2)^2}+{3(6\Delta^2m_\pi^2+14m_\pi^4)\over 
(\Delta^2-m_\pi^2)^{5\over 2}}\ln R\right)\nonumber\\
\delta\beta_{M2}^p&=&-{3\over 5}{L_p\over m_\pi^2}\left({\Delta\over \Delta^2-m_\pi^2}
-{m_\pi^2\over (\Delta^2-m_\pi^2)^{3\over 2}}\ln R\right)\, . 
\end{eqnarray} 
It is interesting to note that
that the sum rules Eq. (\ref{eq:klm}) require both electric polarizabilities to be
positive definite, which is obeyed by the chiral calculation. 

As to the experimental evaluation of such proton structure probes, it is, of
course, in principle possible to extract them directly from Compton cross 
section
measurements, since they modify the scattering amplitude via
\begin{eqnarray}
\delta{\rm Amp}^{(4)}&=&4\pi\omega^4\left[\hat{\epsilon}_2\cdot\hat{\epsilon}_1\left(
\alpha_{E\nu}^p-{1\over 12}\beta_{M2}^p+\hat{k}_2\cdot\hat{k}_1(\beta_{M\nu}^p+
{1\over 12}\alpha_{E2}^p)
+(\hat{k}_2\cdot\hat{k}_1)^2{1\over 6}\beta_{M2}^p\right)\right.\nonumber\\
&+&\left.\hat{\epsilon}_2\cdot\hat{k}_1\hat{\epsilon}_1
\cdot\hat{k}_2\left(-\beta_{M\nu}^p+
{1\over 12}\alpha_{E2}^p-\hat{k}_2\cdot\hat{k}_1{1\over 6}\beta_{M2}^p\right)
\right]\, .
\end{eqnarray}
However, isolating such pieces from other terms which 
affect the cross section at energies above $\sim$ 100 MeV is virtually 
impossible since additional higher order effects soon become equally 
important~\cite{lvov}.
Thus an alternative procedure is required and is made possible by
the validity of dispersion relations, as described in the next section.
 
\section{Dispersive Evaluation}

Assuming the validity of fixed-t dispersion relations it is possible to determine
the quadratic polarizabilities in a relatively model-independent
fashion.  Such a dispersive approach to the calculation of
Compton scattering amplitudes has recently 
been carried out by Drechsel et al.~\cite{dr}.  In this
method one decomposes the center of mass frame 
Compton amplitude in terms of invariant amplitudes
$A_i(\nu,t)$ via
\begin{eqnarray}
{\rm Amp}_{\rm c.m.}&=&-\hat{\epsilon}_1\cdot\hat{\epsilon}_2\omega^2(A_1(1-\cos\theta)+(A_3+A_6)
(1-\cos\theta))\nonumber\\
&+&\hat{\epsilon}_2\cdot
\hat{k}_1\hat{\epsilon}_1\cdot\hat{k}_2\omega^2(A_1+A_3+A_6)\nonumber\\
&+&i\vec{\sigma}\cdot\hat{\epsilon}_2\times\hat{\epsilon}_1{\omega^3\over M}
(-(A_2+A_5)(1-\cos\theta)+A_4(1+\cos\theta))\nonumber\\
&+&i\vec{\sigma}\cdot\hat{k}_2
\times\hat{k}_1\hat{\epsilon}_2\cdot\hat{\epsilon}_1{\omega^3\over M}(A_5-A_6)
\nonumber\\
&+&(\vec{\sigma}\cdot\hat{\epsilon}_2\times\hat{k}_1\hat{\epsilon}_1
\cdot\hat{k}_2-\vec{\sigma}\cdot\hat{\epsilon}_1\times\hat{k}_2
\hat{\epsilon}_2\cdot\hat{k}_1){\omega^3\over 2M}(A_6-2A_5-A_4-A_2)\nonumber\\
&+&(\vec{\sigma}\cdot\hat{\epsilon}_2\times\hat{k}_2\hat{\epsilon}_1
\cdot\hat{k}_1-\vec{\sigma}\cdot\hat{\epsilon}_1\times\hat{k}_1\hat{\epsilon}_2\cdot\hat{k}_2){\omega^2\over 2M}
(A_2-A_4-A_6)\nonumber\\
&+&{\cal O}({\omega^4\over M^2},{\omega^5\over M^3})
\end{eqnarray}  
where $\nu=(s-u)/4M$, $t=-2k_1\cdot k_2$, and assumes that the $A_i$ can be
represented in terms of once subtracted dispersion relations at fixed t
\begin{equation}
\label{schannel-sub}
A_i(\nu,t)=A_i^{\rm Born}(\nu,t)+(A_i(0,t)-A_i^{\rm Born}(0,t))
+{2\nu^2\over \pi}P\int_{\nu_{thr}}^\infty d\nu'{{\rm Im}_sA_i(\nu',t)\over
\nu' ({\nu'}^2-\nu^2)}\, .
\end{equation}
Here Im$_tA_i(\nu',t)$ is evaluated using empirical photoproduction data
while the subtraction constant $A_i(0,t)-A_i^{\rm Born}(0,t)$ is
represented via use of t-channel dispersion relations
\begin{equation}
\label{tchannel-sub}
A_i(0,t)-A_i^{\rm Born}(0,t)=a_i+a_i^{t-pole}+
{t\over \pi}\left(\int_{4 m_\pi^2}^\infty
-\int_{-\infty}^{- 4 M m_\pi - 2 m_\pi^2} dt'
{{\rm Im}_tA_i(0,t')\over t'(t'-t)}\right)
\end{equation}
with Im$_tA_i$ evaluated using the contribution from the $\pi\pi$
intermediate states along the positive-t cut and the s- and u-channel 
$\Delta(1232)$
exchange along the negative-t cut. In principle then there remain six
unknown subtraction constants $a_i$ which can be determined
empirically.  However, in view of the limitations posed by the the data,
Drechsel et al.  note that four of these quantities can be reasonably
assumed to obey unsubtracted forward 
dispersion relations and can be evaluated via
\begin{equation}
a_i={2\over \pi}\int_{\nu_0}^\infty d\nu'{{\rm Im}_sA_i(\nu',t=0)\over \nu'}
\qquad i=3,4,5,6 \, .
\end{equation}
The group treats $a_1,a_2$ as free parameters which can be determined via a
fit to experiment and in this way is able to obtain a very good
description of the low energy Compton scattering data.  

In this note we wish to go a step further and attempt to identify
{\it higher order} terms in the expansion of the Compton amplitudes $A_i$
which can be reasonably evaluated by means of the
{\it subtracted} dispersion relations.
Defining $a_{i,t}, a_{i,\nu}$ as the appropriate derivatives at
$t,\nu^2=0$, i.e.
\begin{eqnarray}
a_{i,\nu}&=&{2\over \pi}\int_{\nu_{thr}}^\infty d\nu'{{\rm
    Im}_sA_i(\nu',t=0)\over \nu'^3},\\
a_{i,t}&=& {1\over \pi}\left(\int_{4 m_\pi^2}^\infty
-\int_{-\infty}^{- 4 M m_\pi - 2 m_\pi^2} dt'
{{\rm Im}_tA_i(0,t')\over t'^2}\right),
\end{eqnarray}
we have~\cite{rev}\footnote{Here the small nonderivative 
recoil terms arise from the transformation from the Breit frame,
wherein the effective Hamiltonian description is defined, to the
center of mass frame, in which we work\cite{rev}. }
\begin{eqnarray}
4\pi\alpha_{E2}^p&=&-12(a_{3,t}+a_{6,t}+a_{1,t})+{3\over M^2}a_3\nonumber\\
4\pi\beta_{M2}^p&=&-12(a_{3,t}+a_{6,t}-a_{1,t})+{3\over M^2}a_3\nonumber\\
4\pi\alpha_{E\nu}^p&=&-a_{3,\nu}-a_{6,\nu}-a_{1,\nu}+a_{3,t}+a_{6,t}+3a_{1,t}
-{1\over 4M^2}(a_3+4a_5)\nonumber\\
4\pi\beta_{M\nu}^p&=&-a_{3,\nu}-a_{6,\nu}+a_{1,\nu}+a_{3,t}+a_{6,t}-3a_{1,t}
-{1\over 4M^2}(a_3-4a_5),
\end{eqnarray} 
and in this way one finds the values (all in units of $10^{-4}$ fm$^5$)
\begin{equation}
{\rm DR}:
\quad\alpha_{E\nu}^p=-3.84,
\quad\beta_{M\nu}^p=9.29,\quad
\alpha_{E2}^p=29.31,\quad
\beta_{M2}^p=-24.33\, .
\end{equation}
These numbers are in reasonably good agreement with those obtained in Ref. 
\cite{rev} which assumes
the validity of an unsubtracted dispersion relation and append
high energy behavior
in the cross channel,
\begin{equation}
{\rm DR}:\cite{rev}\quad\alpha_{E\nu}^p=-3.8,\quad\beta_{M\nu}^p=9.1,\quad\alpha_{E2}^p=27.5,
\quad\beta_{M2}^p=-22.4 \, .
\end{equation}
The (relatively small) differences between the two evaluations should perhaps be considered
as an indication of the uncertainty in the extraction.

Now we can confront these values with the corresponding 
theoretical calculations.
In the case of the chiral predictions at ${\cal O}(p^3)$ we have from 
Eq.~(\ref{eq:cde})
\begin{equation}
{\cal O}(p^3):\quad\alpha_{E\nu}^p=2.4,\quad\beta_{M\nu}^p=3.7,\quad
\alpha_{E2}^p=22.1,\quad\beta_{M2}^p=-9.5 \, .
\end{equation}
We see then that the size of $\alpha_{E2}^p$ is about right, while for
both $\beta_{M2}^p$ and $\beta_{M\nu}^p$ the sign and order of magnitude
is correct but additional contributions are called for.  The most
serious problem lies in the experimental determination of
$\alpha_{E\nu}$ which is negative in contradistinction to the chiral
prediction and to sum rule arguments which assert its positivity.  Of
course, the experimental (and theoretical) numbers are small so
perhaps the disagreement lies within the uncertainty of our
evaluation.  Equivalently it could be that a nonrelativistic constituent quark
model approach to subtle details of proton structure is inappropriate.  These
issues should be addressed in future work.

We can now move on to consider whether corrections from 
$\Delta(1232)$ degrees of freedom can help to address any
discrepancies found above.   We find
\begin{equation}
{\cal O}(\epsilon^3):\quad \alpha_{E\nu}^p=1.7,\quad\beta_{M\nu}^p=7.5,\quad
\alpha_{E2}^p=26.2,\quad\beta_{M2}^p=-12.3 \, .
\end{equation}
Except for the sign problem with $\alpha_{E\nu}^p$ indicated above, which remains in the 
$\epsilon$-expansion, the changes are generally helpful, although the magnetic 
quadrupole polarizability is still somewhat underpredicted.

\section{Higher Order Spin Polarizabilities}

One can also analyze higher order contributions to spin 
polarizabilities.  In this case the leading 
order---${\cal O}(\omega^3)$---effective Lagrangian reads
\begin{equation}
H_{eff}^{(3)}=-{1\over 2}4\pi(\gamma_{E1}^p\vec{\sigma}\cdot\vec{E}\times
\dot{\vec{E}}+\gamma_{M1}^p\vec{\sigma}\cdot\vec{H}\times
\dot{\vec{H}}-2\gamma_{E2}^pE_{ij}\sigma_iH_j+2\gamma_{M2}^pH_{ij}\sigma_iE_j)
\label{eq:mno}
\end{equation} 
and the chiral predictions for the spin-polarizabilities at ${\cal O}(p^3)$ 
are found to be\cite{hhk}
\begin{equation}
\gamma_{E1}^p=-{10K_p\over \pi m_\pi},\quad\gamma_{M1}^p=-{2K_p\over \pi 
m_\pi},\quad\gamma_{E2}^p=
{2K_p\over \pi m_\pi},\quad\gamma_{M2}^p={2K_p\over \pi m_\pi} \, .
\end{equation}
Numerical values 
are given below in units of $10^{-4}$ fm$^4$
\begin{equation}
{\cal O}(p^3):\quad\gamma_{E1}^p=-5.8,\quad\gamma_{M1}^p
=-1.2,\quad\gamma_{E2}^p
=1.2,\quad\gamma_{M2}^p=1.2 \, .\label{eq:tuv}
\end{equation}
It is interesting to note that 
$\gamma_{E1}$ is nearly an order of magnitude larger than the other 
spin-polarizabilities.\footnote{Note that before comparison with experiment 
is made these terms must be augmented by contributions from the ``anomaly'' 
({\it i.e } pion pole graph)---
\begin{equation}
{}_a\gamma_{E1}^p=-{}_a\gamma_{M1}^p=-{}_a\gamma_{E2}^p={}_a\gamma_{M2}^p
={24K_p\over \pi m_\pi g_A }\, .
\end{equation}}

Thusfar, experiments utilizing a polarized target and beam, which are necessary
in order to directly measure the spin-polarizabilities, have not been 
performed.  However, one can compare with dispersion relation predictions, as
done above.  Since each involves spin-flip amplitudes, unsubtracted  
integrals are expected to converge and one finds values
\begin{equation}
{\rm DR}:\quad\gamma_{E1}^p=-4.5,\quad\gamma_{M1}^p=3.4,\quad\gamma_{E2}^p
=2.3,\quad\gamma_{M2}^p=-0.6
\end{equation}
which are in reasonable agreement with the numbers
\begin{equation}
{\rm DR}:\cite{rev}\quad\gamma_{E1}^p=-3.4,\quad\gamma_{M1}^p=2.7,\quad
\gamma_{E2}^p=1.9,\quad\gamma_{M2}^p=0.3
\end{equation}
extracted in Ref. \cite{rev}.  Again the sign discrepancy in the small term
$\gamma_{M2}^p$ is perhaps an indication of the overall precision which one
can expect via the dispersive procedure.  In comparing with the chiral 
numbers---Eq. (\ref{eq:tuv})---we observe that the predictions for both electric 
multipoles are quite satisfactory.  However, there is a clear problem in the 
comparison for $\gamma_{M1}^p$, suggesting the necessity of including the 
contributions from the $\Delta(1232)$, which are found to be
\begin{eqnarray}
\delta\gamma_{E1}^p&=& {L_p\over 12}\left({\Delta^2+5m_\pi^2\over \Delta^2-m_\pi^2}\right)\left({1\over \Delta^2-m_\pi^2}-{\Delta\over (\Delta^2-m_\pi^2)^{3\over 2}}\ln R\right)\nonumber\\
\delta\gamma_{M1}^p&=&{4\over 9}{b_1^2\alpha\over M^2\Delta^2}-{L_p\over 12}\left({1\over 
\Delta^2-m_\pi^2}-{\Delta\over (\Delta^2-m_\pi^2)^{3\over 2}}\ln R\right)\nonumber\\
\delta\gamma_{E2}^p&=&\delta\gamma_{M2}^p={L_p\over 12}\left({1\over \Delta^2-m_\pi^2}-
{\Delta\over (\Delta^2-m_\pi^2)^{3\over 2}}\ln R\right)\, .
\end{eqnarray}
There does exist then a significant contribution to $\gamma_{M1}^p$ from the 
$\Delta(1232)$ pole diagram as well as small contributions to the other
spin-polarizabilities from $\Delta\pi$ loop effects.  When these are appended to
the $N\pi$ loop predictions given in Eq. (\ref{eq:tuv}) we find results
\begin{equation}
{\cal O}(\epsilon^3):\quad\gamma_{E1}^p=-6.1,\quad\gamma_{M1}^p=1.0,
\quad\gamma_{E2}^p=1.1,\quad\gamma_{M2}^p=1.1
\end{equation}
which are in quite reasonable agreement with the numbers obtained dispersively
above.

However, it is also possible to study {\it higher order} polarizability 
contributions to the spin-dependent 
Compton scattering amplitude, which contribute at ${\cal O}(\omega^5)$.  There
are eight such new terms, which can be expressed in terms of the effective
Hamiltonian
\begin{eqnarray}
H_{eff}^{(5)}&=&-{1\over 2}4\pi\left[\gamma_{E1\nu}^p\vec{\sigma}\cdot\dot{\vec{E}}
\times\ddot{\vec{E}}+\gamma_{M1\nu}^p\vec{\sigma}\cdot\dot{\vec{H}}\times
\ddot{\vec{H}}-2\gamma_{E2\nu}^p\sigma_i\dot{E}_{ij}\dot{H}_j+2\gamma_{M2\nu}^p
\sigma_i\dot{H}_{ij}\dot{E_j}\right.\nonumber\\
&+&\left.4\gamma_{ET}^p\epsilon_{ijk}\sigma_iE_{j\ell}\dot{E}_{k\ell}
+4\gamma_{MT}^p\epsilon_{ijk}\sigma_iH_{j\ell}\dot{H}_{k\ell}
-6\gamma_{E3}^p\sigma_iE_{ijk}H_{jk}+6\gamma_{M3}^p\sigma_iH_{ijk}E_{jk}
\right]\nonumber\\
\quad 
\end{eqnarray}
where 
\begin{eqnarray}
{(E,H)}_{ijk}&=&{1\over 3}(\nabla_i\nabla_j{(E,H)}_k+\nabla_i\nabla_k{(E,H)}_j
+\nabla_j\nabla_k{(E,H)}_i)\nonumber\\
&-&{1\over
15}(\delta_{ij}\nabla^2(E,H)_k+\delta_{jk}
\nabla^2(E,H)_i+\delta_{ik}\nabla^2(E,H)_j)
\end{eqnarray}
are the (spherical) tensor gradients of the electric and magnetizing 
fields.\footnote{For completeness, we
note that these higher order spin-polarizabilities can be expressed, neglecting recoil terms, in
terms of the usual multipole expansion via
\begin{eqnarray}
4\pi\omega^5\gamma_{ET}&=&3(f_{EE}^{2+}-f_{EE}^{2-})\nonumber\\
4\pi\omega^5\gamma_{MT}&=&3(f_{MM}^{2+}-f_{MM}^{2-})\nonumber\\
4\pi\omega^5\gamma_{E3}&=&15f_{ME}^{2+}\nonumber\\
4\pi\omega^5\gamma_{M3}&=&15f_{EM}^{2+}\nonumber\\
4\pi\omega^3(\gamma_{E2}+\omega^2\gamma_{E2\nu})&=&6f_{ME}^{1+}\nonumber\\
4\pi\omega^3(\gamma_{M2}+\omega^2\gamma_{M2\nu})&=&6f_{EM}^{1+}\nonumber\\
4\pi\omega^3(\gamma_{E1}+\omega^2\gamma_{E1\nu})&=&
f_{EE}^{1+}-f_{EE}^{1-}\nonumber\\
4\pi\omega^3(\gamma_{M1}+\omega^2\gamma_{M1\nu})&=&
f_{MM}^{1+}-f_{MM}^{1-}\, .
\end{eqnarray}}
We see that, as in the spin-averaged case, four of the new terms are simply 
dispersive corrections to the ${\cal O}(\omega^3)$ spin-polarizabilities 
defined in Eq. (\ref{eq:mno}).  However, there exist also new structures which
probe the octupole excitation of the system.  The modification of the Compton
scattering amplitude by such terms is found to be
\begin{eqnarray}
\delta{\rm Amp}^{(5)}&=&-i4\pi\omega^5[\vec{\sigma}\cdot\hat{\epsilon}_2\times
\hat{\epsilon}_1(\gamma_{E1\nu}^p+\gamma_{M2\nu}^p-{12\over 5}\gamma_{E3}^p-
3\gamma_{MT}^p\nonumber\\
&+&\hat{k}_2\cdot\hat{k}_1(\gamma_{M1\nu}^p+\gamma_{E2\nu}^p+\gamma_{ET}^p
+{8\over 5}\gamma_{M3}^p)+4(\hat{k}_2\cdot\hat{k}_1)^2(\gamma_{MT}^p+\gamma_{E3}^p))\nonumber\\
&+&\vec{\sigma}\cdot\hat{k}_2 \times\hat{k}_1\hat{\epsilon}_2
\cdot\hat{\epsilon}_1
(\gamma_{M1\nu}^p-\gamma_{E2\nu}^p+\gamma_{ET}^p
+{2\over 5}\gamma_{M3}^p+\hat{k}_2\cdot\hat{k}_1
(4\gamma_{MT}^p-2\gamma_{E3}^p))
\nonumber\\
&+&(\vec{\sigma}\cdot\hat{\epsilon}_2\times\hat{k}_1\hat{\epsilon}_1
\cdot\hat{k}_2
-\vec{\sigma}\cdot\hat{\epsilon}_1\times\hat{k}_2\hat{\epsilon}_2\cdot
\hat{k}_1)(\gamma_{ET}^p+\gamma_{M3}^p-\gamma_{M1\nu}^p-\hat{k}_2\cdot\hat{k}_1(4\gamma_{MT}^p+\gamma_{E3}^p))
\nonumber\\
&+&(\vec{\sigma}\cdot\hat{\epsilon}_2\times\hat{k}_2\hat{\epsilon}_1
\cdot\hat{k}_2
-\vec{\sigma}\cdot\hat{\epsilon}_1\times\hat{k}_1
\hat{\epsilon}_2\cdot\hat{k}_1)({7\over 5}\gamma_{E3}^p+
2\gamma_{MT}^p-\gamma_{M2\nu}^p-3\hat{k}_2\cdot\hat{k}_1
\gamma_{M3}^p)]
\end{eqnarray}
and, comparing with a chiral expansion of the Compton amplitude at 
${\cal O}(p^3)$, we can read off\cite{gmr}\footnote{As before, there exist 
pion pole 
(anomaly) contributions to the higher order spin polarizabilities which must 
be included when comparing with data
\begin{equation}
\begin{array}{cccc}
{}_a\gamma_{E1\nu}^p=-{72K_p\over \pi g_Am_\pi^3}&
{}_a\gamma_{M1\nu}^p={72K_p\over \pi g_Am_\pi^3}&
{}_a\gamma_{E2\nu}^p={368K_p\over 5\pi g_Am_\pi^3}&
{}_a\gamma_{M2\nu}^p=-{368K_p\over 5\pi g_Am_\pi^3}\\
{}_a\gamma_{E3}^p=-{16K_p\over \pi g_Am_\pi^3}&
{}_a\gamma_{M3}^p={16K_p\over \pi g_Am_\pi^3}&
{}_a\gamma_{ET}^p={8K_p\over \pi g_Am_\pi^3}&
{}_a\gamma_{MT}^p=-{8K_p\over \pi g_Am_\pi^3}\, .
\end{array}
\end{equation}}
\begin{eqnarray}
\gamma_{E3}^p&=&{4K_p\over 45\pi m_\pi^3},\quad\gamma_{M3}^p
={4K_p\over 45\pi m_\pi^3}
,\quad
\gamma_{ET}^p=-{13K_p\over 45\pi m_\pi^3},\quad
\gamma_{MT}^p=-{K_p\over 45\pi m_\pi^3}
\nonumber\\
\gamma_{E1\nu}^p&=&-{189K_p\over 45\pi m_\pi^3},\quad\gamma_{M1\nu}^p=-
{9K_p\over 45\pi m_\pi^3}
,\quad
\gamma_{E2\nu}^p={78K_p\over 225\pi m_\pi^3},\quad\gamma_{M2\nu}^p
=-{42K_p\over 225\pi m_\pi^3}\, .\nonumber\\
\quad 
\end{eqnarray}
We have then the chiral predictions (in units of $10^{-4}$ fm$^6$) 
\begin{equation}
\begin{array}{ccccc}
{\cal O}(p^3):&
\gamma_{E3}^p=0.11&
\gamma_{M3}^p=0.11&
\gamma_{ET}^p=-0.37&
\gamma_{MT}^p=-0.03\\
\quad &
\gamma_{E1\nu}^p=-5.05&
\gamma_{M1\nu}^p=-0.26&
\gamma_{E2\nu}^p=0.45&
\gamma_{M2\nu}^p=-0.24\, .
\end{array}
\end{equation}
Again we note that the size of $\gamma_{E1\nu}^p$ dominates by 
over an order of magnitude any of
the other higher order spin-polarizabilities.  
The modifications arising from inclusion of the $\Delta(1232)$ are found to be
\begin{eqnarray}
\delta\gamma_{E3}^p&=&-{L_p\over 540}\left({\Delta^2
+2m_\pi^2\over m_\pi^2(\Delta^2-m_\pi^2)^2}
-{3\Delta\over (\Delta^2-m_\pi^2)^{5/2}}\ln R\right)\nonumber\\
\delta\gamma_{M3}^p&=&-{L_p\over 540}\left({\Delta^2+2m_\pi^2\over m_\pi^2
(\Delta^2-m_\pi^2)^2}-{3\Delta\over (\Delta^2-m_\pi^2)^{5\over 2}}\ln R\right)\nonumber\\
\delta\gamma_{ET}^p&=&{L_p\over 1080}\left({2\Delta^4+50\Delta^2m_\pi^2-13m_\pi^4
\over m_\pi^2(\Delta^2-m_\pi^2)^3}-{3\Delta({1\over 2}\Delta^2-8m_\pi^2)\over 
(\Delta^2-m_\pi^2)^{7\over 2}}\ln R\right)\nonumber\\
\delta\gamma_{MT}^p&=&{L_p\over 2160}\left({\Delta^2+2m_\pi^2\over 
m_\pi^2(\Delta^2-m_\pi^2)^2}-{3\Delta\over (\Delta^2-m_\pi^2)^{5\over 2}}
\ln R\right)\nonumber\\
\delta\gamma_{E1\nu}^p&=&{L_p\over 720}\left({-5\Delta^6+26\Delta^4m_\pi^2+
693\Delta^2m_\pi^4+126m_\pi^6\over m_\pi^2(\Delta^2-m_\pi^2)^4}\right.\nonumber\\
&-&\left.{3\Delta(4\Delta^4+152\Delta^2m_\pi^2+124m_\pi^4)
\over (\Delta^2-m_\pi^2)^{9\over 2}}\ln R\right)\nonumber\\
\delta\gamma_{M1\nu}^p&=&{4\over 9}{b_1^2\alpha\over M^2\Delta^4}+
{L_p\over 1080}\left({-3\Delta^4+9\Delta^2m_\pi^2-9m_\pi^4\over
m_\pi^2(\Delta^2-m_\pi^2)^3}+{27\Delta({3\over 2}\Delta^2+m_\pi^2)\over 
(\Delta^2-m_\pi^2)^{7\over 2}}\ln R\right)\nonumber\\
\delta\gamma_{E2\nu}^p&=&{L_p\over
1800}\left({7\Delta^4-93\Delta^2m_\pi^2
+26m_\pi^4\over m_\pi^2(\Delta^2-m_\pi^2)^3}-
{3\Delta(27\Delta^2+23m_\pi^2)\over (\Delta^2-m_\pi^2)^{7\over 2}}
\ln R\right)\nonumber\\
\delta\gamma_{M2\nu}^p&=&{L_p\over
1800}\left({17\Delta^4+72\Delta^2m_\pi^2
-14m_\pi^4\over
m_\pi^2(\Delta^2-m_\pi^2)^3}-{3\Delta(27\Delta^2+2m_\pi^2)
\over (\Delta^2-m_\pi^2)^{7\over 2}}
\ln R\right)
\end{eqnarray}
and numerically this leads to the predictions
\begin{eqnarray}
\begin{array}{ccccc}
{\cal O}(\epsilon^3):&
\gamma_{E3}^p=0.11,&
\gamma_{M3}^p=0.11,&
\gamma_{ET}^p=-0.28,&
\gamma_{MT}^p=-0.03\\
&\gamma_{E1\nu}^p=-5.16,&
\gamma_{M1\nu}^p=0.83,&
\gamma_{E2\nu}^p=0.28,
&\gamma_{M2\nu}^p=-0.22\, .
\end{array}\nonumber\\
\quad
\end{eqnarray}
The higher order polarizabilities can be extracted from the dispersive results
via the relations
\begin{eqnarray}
4\pi\gamma_{E3}^p&=&-{1\over 3M}(a_{4,t}+a_{6,t}+a_{2,t})+{1\over
12M^3}a_5\nonumber\\
4\pi\gamma_{M3}^p&=&-{1\over 3M}(a_{4,t}+a_{6,t}-a_{2,t})-{1\over
12M^3}a_5\nonumber\\
4\pi\gamma_{ET}^p&=&{1\over 6M}(2a_{6,t}+3a_{5,t}-a_{4,t}+a_{2,t})
+{1\over 48M^3}(-6a_3+7a_5+3a_6)\nonumber\\
4\pi\gamma_{MT}^p&=&{1\over 6M}(2a_{6,t}-3a_{5,t}-a_{4,t}-a_{2,t})
+{1\over 48M^3}(-6a_3-7a_5+3a_6)\nonumber\\
4\pi\gamma_{E2\nu}^p&=&-{1\over 2M}(a_{4,\nu}+a_{6,\nu}+a_{2,\nu})
+{1\over 5M}(6a_{6,t}+5a_{5,t}+a_{4,t}+9a_{2,t})\nonumber\\
&-&{1\over 40M^3}(-10a_3+90a_4-7a_5+15a_6)\nonumber\\
4\pi\gamma_{M2\nu}^p&=&-{1\over 2M}(a_{4,\nu}+a_{6,\nu}-a_{2,\nu})
+{1\over 5M}(6a_{6,t}-5a_{5,t}+a_{4,t}-9a_{2,t})\nonumber\\
&-&{1\over 40M^3}(-10a_3+90a_4+7a_5+15a_6)\nonumber\\
4\pi\gamma_{E1\nu}^p&=&{1\over 2M}(a_{6,\nu}+2a_{5,\nu}-a_{4,\nu}+a_{2,\nu})
+{1\over 2M}(-2a_{6,t}-5a_{5,t}+a_{4,t}-3a_{2,t})\nonumber\\
&+&{1\over 16M^3}(-2a_3-36a_4+19a_5+5a_6)\nonumber\\
4\pi\gamma_{M1\nu}^p&=&{1\over 2M}(a_{6,\nu}-2a_{5,\nu}-a_{4,\nu}-a_{2,\nu})
+{1\over 2M}(-2a_{6,t}+5a_{5,t}+a_{4,t}+3a_{2,t})\nonumber\\
&+&{1\over 16M^3}(-2a_3-36a_4-19a_5+5a_6)
\end{eqnarray}
which yields
\begin{eqnarray}
DR:
\quad\gamma_{E3}^p&=&0.06,
\quad\gamma_{M3}^p=0.09,
\quad\gamma_{ET}^p=-0.15,
\quad\gamma_{MT}^p=-0.09\nonumber\\
\gamma_{E1\nu}^p&=&-2.84,
\quad\gamma_{M1\nu}^p=2.23,
\quad\gamma_{E2\nu}^p=1.03,
\quad\gamma_{M2\nu}^p=-0.60  \, . \nonumber \\
\end{eqnarray}
Obviously only the dispersive correction coefficients $\gamma_{E1\nu},
\gamma_{M1\nu},\gamma_{E2\nu},\gamma_{M2\nu}$ are sizable and are in
qualitative agreement with the chiral ${\cal O}(\epsilon^3)$ predictions.
 
\section{Conclusions}

Above we have shown how the use of dispersion relations allows extraction
of information about higher order polarizabilities of the proton which is
not available from direct cross section analysis.  We have also seen how such 
measurements can be confronted with theoretical predictions for such quantities
based on quark model and/or chiral perturbative pictures of proton structure.
Although a simple harmonic oscillator model contains too small a gap between
the ground and excited states and therefore overpredicts both the conventional
as well as the higher order polarizabilities, a simple ${\cal O}(p^3)$ or
${\cal O}(\epsilon)^3$ HB$\chi$PT is in basic agreement with the dispersive
evaluation, except for a sign problem in the case of $\alpha_{E\nu}$.  Since
general sum rule arguments disagree with the sign of the experimentally 
extracted term, this is clearly an area which demands additional study.
We also presented theoretical predictions for higher 
order---${\cal O}(\omega^5)$---contributions to the spin-polarizabilities, 
which can 
in principle be extracted once spin-dependent data become available.   
Clearly there is a great deal of nucleon structure information contained in
such higher order polarizabilities and our paper has just touched the surface.

\begin{center}
{\bf Acknowledgement}
\end{center}

BH would like to acknowledge the support of the Alexander von Humboldt 
Foundation as well as the hospitality of Forschungszentrum J\"{u}lich.  This 
work was also supported in part by the National Science Foundation and
by the Deutsche Forschungsgemeinschaft (SFB 443).


\medskip

\begin{thebibliography}{99}

\bibitem{Nat96} F.J. Federspiel et al., Phys. Rev. Lett. {\bf 67},
1511 (1991); E.L. Hallin et al., Phys. Rev. {\bf C48}, 1497 (1993); A.
Zieger et al., Phys. Lett. {\bf B278}, 34 (1992); B.E. MacGibbon et
al., Phys. Rev. {\bf C52}, 2097 (1995).
\bibitem{pedag}  See, {\it e.g.}, B.R. Holstein, Am. J. Phys. {\bf 67},
422 (1999).  
\bibitem{bal} A.M. Baldin, Nucl. Phys. {\bf 18}, 310 (1960); L.I.
Lapidus, Sov. Phys. JETP {\bf 16}, 964 (1963).
\bibitem{Damashek}
M. Damashek and F.J. Gilman, Phys. Rev. D {\bf 1}, 1319 (1970).
\bibitem{BGM}
D. Babusci, G. Giordano and G. Matone, Phys. Rev. C {\bf 57}, 291 (1998).
\bibitem{merz} See, {\it e.g.}, E. Merzbacher, {\it Quantum Mechanics},
Wiley, New York (1998), Ch. 18.4.
\bibitem{Fr75} See, {\it e.g.}, D.H. Perkins, {\it Introduction to High
Energy Physics}, Addison-Wesley, Reading, MA (1987).
\bibitem{ww} R. Weiner and W. Weise, Phys. Lett. {\bf B159}, 85 (1985). 
\bibitem{BKM95} V. Bernard, N. Kaiser, and U.-G. Meissner, Int. J.
Mod. Phys. {\bf E4}, 193 (1995).
\bibitem{BKKM92} V. Bernard, N. Kaiser, J. Kambor, and U.-G. Meissner,
Nucl. Phys. {\bf B388}, 315 (1992). 
\bibitem{BKSM93} V. Bernard, N. Kaiser, A. Schmidt, and U.-G. Meissner,
Phys. Lett. {\bf B319}, 269 (1993): Z. Phys. {\bf A348}, 317 (1994).
\bibitem{hem} T.R. Hemmert, B.R. Holstein, and J. Kambor, J. Phys.
{\bf G24}, 1831 (1998).
\bibitem{gmr1} T.R. Hemmert, B.R. Holstein, and J. Kambor, Phys. Rev.
{\bf D55}, 5598 (1997); 
see also M.N. Butler and M.J. Savage, Phys. Lett. B {\bf 294} (1992) 369.
\bibitem{gmr} G. Kn\"{o}chlein, Universit\"{a}t Mainz Ph.D. Thesis, Shaker
  Verlag (Aachen) 1997. 
\bibitem{M9} N. Mukhopadhyay, A.M. Nathan, and L. Zhang, Phys. Rev.
{\bf D47}, R7 (1993).
\bibitem{lvov} A.I. L'vov, Int. J. Mod. Phys. {\bf A8}, 5267 (1993).
\bibitem{rev} D. Babusci, G. Giordano, A.I. L'vov, G.
Matone, and A.M. Nathan, Phys. Rev. {\bf C58}, 1013 (1998).
\bibitem{dr} D. Drechsel, M. Gorchtein, B. Pasquini, and M.
Vanderhaeghen, hep-ph/9904290; Phys. Rev. C in press.
\bibitem{hhk} T.R. Hemmert, B.R. Holstein, J. Kambor, and
  G. Kn\"{o}chlein, Phys. Rev. {\bf D57}, 5746 (1998).

\end{thebibliography}
\end{document}